\documentclass[aps,prl,twocolumn,showpacs,preprintnumbers,a4paper]{revtex4} 
\usepackage{graphics}
\usepackage{epsf}
\usepackage{epsfig}
\usepackage{amsmath}
\usepackage{amsfonts}
\usepackage{amssymb}
\usepackage{graphicx}%
\begin{document}
\title{Finite-size scaling of directed percolation above the upper critical dimension}
\author{S.~L\"ubeck}
\affiliation{Institut f\"ur Theoretische Physik, Universit\"at-Duisburg-Essen, 47048
Duisburg, Germany}
\author{H.-K.~Janssen}
\affiliation{Institut f\"ur Theoretische Physik III, Heinrich-Heine-Universit\"at, 40225
D\"usseldorf, Germany}
\date{\today}

\begin{abstract}
We consider analytically as well as numerically the finite-size scaling
behavior in the stationary state near the non-equilibrium phase
transition of directed percolation within the mean field regime, i.e., above
the upper critical dimension. Analogous to equilibrium, usual finite-size
scaling is valid below the upper critical dimension, whereas it fails above.
Performing a momentum analysis of associated path integrals we derive modified
finite-size scaling forms of the order parameter and its higher moments.
The results are confirmed by
numerical simulations of corresponding high-dimensional lattice models.

\end{abstract}
\maketitle



\preprint{{\it Physical Review E}{\textbf 72}, 016119 2010}

Critical phenomena such as second order phase transitions are characterized by
singularities causing a discontinuous behavior of various quantities at the
transition point (e.g.~the specific heat, the susceptibility, 
and the correlation length).
These singularities are described by power-laws defining the well-known
critical exponents. Studying the phase transition of a given system, the aim
of investigation is to identify the set of critical exponents which
characterizes (together with certain universal scaling functions) the
so-called universality class. Since most systems are not analytically
tractable their critical behavior is often investigated via 
numerical methods, for example Monte Carlo simulations or transfer 
matrix calculations. 
In these
cases, the obtained data are limited to finite system sizes. Therefore,
finite-size scaling (FSS) is widely used to extrapolate to the behavior of the
infinite systems. In particular, FSS is an efficient method to determine the
critical exponents and provides certain universal scaling functions, i.e., it
allows to identify the universality class (see~\cite{BARBER_1,Ca88} for reviews).

According to the phenomenological FSS theory~\cite{FISHER_7}, a finite system
size~$L$ results in a rounding and shifting of the singularities. Furthermore,
it is assumed that finite-size effects are controlled sufficiently close to
the critical point by the ratio $L/\xi_{\infty}$. Here, $\xi_{\infty}$ denotes
the spatial correlation length of the infinite system. Approaching the
transition point, the correlation length diverges as $\xi_{\infty}%
\propto\left\vert \tau-\tau_{c}\right\vert ^{-\nu}$, with the critical
exponent~$\nu$ and where $\tau$ is a temperature-like variable which describes
the distance to the critical point. Finite-size effects are negligible for
$L\gg\xi_{\infty}$ (i.e.~$L\,\left\vert \tau-\tau_{c}\right\vert ^{\nu}\gg1$).
Otherwise, they are relevant, i.e., rounding and shifting effects
occur if $L<\xi_{\infty}$. It is known in equilibrium that the hypothesis of
the fundamental role of the ratio $L/\xi_{\infty}$ is valid only below the
so-called upper critical dimension~$d_{\mathrm{c}}$ (see~\cite{BRANKOV_2} for
a recent review). Above this marginal dimension, mean field theories provide
exact values of the critical exponents as well as of the scaling functions.
Renormalization group treatments show that the failure of usual FSS within the
mean field regime is related to variables (scaling fields) which
become dangerously irrelevant for $d>d_{\mathrm{c}}$~\cite{FISHER_5}.
Dangerous irrelevant variables affect the scaling behavior qualitatively and
furthermore cause the breakdown of hyperscaling laws which connect the
critical exponents to the spatial dimension~$d$. 
Investigations of this breakdown of usual finite size scaling 
trace back to~\cite{BREZIN_5}. After controversial discussions
(see e.g.~\cite{ISING_DILEMMA} and reference therein) the problem
was recently resolved~\cite{CHEN_3} and a convincing agreement between 
numerical data and field theoretical results is 
achieved~\cite{CHEN_3,STAUFFER_5}.

Compared to the equilibrium situation less is known in the case of
non-equilibrium phase transitions. Therefore, we consider in this work the
absorbing phase transition of directed percolation (DP) as an exemplification.
According to its robustness and ubiquity (including critical phenomena in
physics, biology, epidemiology, as well as catalytic chemical
reactions) DP is recognized as the paradigm of non-equilibrium phase
transitions into absorbing states (see~\cite{HINRICHSEN_1} for a readable
review).
Although an analytical solution is still lacking, 
DP plays a comparable role in the realm of non-equilibrium phase
transitions as the Ising model in equilibrium. 
Previous investigations of FSS of DP
focus to the absorbing phase below~$d_{\mathrm{c}}$, 
where $d_{\mathrm{c}}=4$~\cite{JANSSEN_9}.
Here, we are interested in finite-size properties above~$d_{\mathrm{c}}$.
In particular, we study the steady state scaling behavior
of finite systems
in the active phase which is maintained by a homogeneous source.
Using a momentum space analysis of path integrals associated
to the field theoretical formulation of DP, we derive FSS
exponents and universal scaling functions.
Analogous to equilibrium, we demonstrate that usual FSS
has to be modified in order to describe the scaling behavior within the mean
field regime. 
Additional numerical simulations confirm the field theoretical results.
But in contrast to equilibrium 
we observe a convincing agreement between the lowest mode 
finite-size analysis and corresponding numerical results.

The asymptotic behavior of the DP universality class is
described by a minimal stochastic Markovian process represented by the
Langevin equation~\cite{JANSSEN_1,Ja01}
\begin{equation}
\lambda^{-1}{\partial_{t} n }=
-\Big(\tau+\frac{g}{2}\, n-\nabla^{2}\Big)n
+h+\zeta .
\label{eq:langevin_dp_01}
\end{equation}
Here, the density of an active agent $n(\mathbf{r},t)$,
defined on a mesoscopic (coarse grained) scale, corresponds to the
order parameter of the non-equilibrium phase transition. The control parameter
of the transition $\tau$ attains its critical value in an infinite volume
at $\tau_{c}$. 
The homogeneous source~$h$ is conjugated to the order parameter and is usually
implemented as a spontaneous creation of activity (see e.g.~\cite{LUEB_27}). 
For zero $h$, a finite positive density occurs above the
transition point ($\tau<\tau_{c}$) whereas the absorbing vacuum state ($n=0$)
is approached below the transition point.
Furthermore, $\zeta(\mathbf{r},t)$ denotes the noise which accounts
for fluctuations of the density $n(\mathbf{r},t)$. 
This zero-mean Gaussian noise represents fast degrees of freedom 
which were eliminated by a suitable coarse graining procedure.
The noise correlator
\begin{equation}
\overline{\zeta(\mathbf{r},t)\zeta(\mathbf{r}^{\prime},t^{\prime})}
=\lambda^{-1}\, g^{\prime}\,n(\mathbf{r},t)
\,\delta(\mathbf{r}-\mathbf{r}^{\prime})\,
\delta(t-t^{\prime})
\label{eq:langevin_dp_corr_01}
\end{equation}
is dictated by the existence of the absorbing state $n=0$.

Renormalization group techniques have been applied to determine the critical
exponents and the universal scaling
functions~\cite{OBUKHOV_2,JANSSEN_1,Ja01,CARDY_1,JANSSEN_9,JANSSEN_2}. In that case path
integral formulations are more adequate than the Langevin
equation approach~\cite{JANSSEN_8,DEDOMINCIS_1}. Stationary correlation functions
as well as response functions can be determined by calculating path integrals
with weight $\exp{(-\mathcal{J})}$, where the dynamic functional $\mathcal{J}$
describes the considered stochastic process. 
The following dynamic response functional~\cite{JANSSEN_1,Ja01} 
\begin{align}
\mathcal{J}  &  =\int d^{d}r\,dt\,{\lambda}\Big\{\tilde{n}\Big(
\,\lambda^{-1}\partial_{t}+(\tau-\nabla^{2})
\label{eq:action_reggeon_field_theory}\\
&  +\frac{g}{2}\left(  n-\tilde{n}\right)  \Big)n-h\tilde{n}\Big\}\nonumber
\end{align}
is associated to the stochastic process defined by 
Eq.\,(\ref{eq:langevin_dp_01}) and Eq.\,(\ref{eq:langevin_dp_corr_01}).
Here, $\tilde{n}(\mathbf{r},t)$ denotes the response field conjugated to the
Langevin noise.
Furthermore, the coupling constants $g$ and $g^{\prime}$ 
are equated by an appropriate rescaling with the redundant parameter~$K$
\begin{equation}
\tilde{n}(\mathbf{r},t)\rightarrow K^{-1}\tilde{n}(\mathbf{r},t)\,,\,
n(\mathbf{r},t)\rightarrow Kn(\mathbf{r},t)\,,\; h\rightarrow K h .
\label{eq:rescal_field}%
\end{equation}
The functional $\mathcal{J}$ is invariant under the time inversion 
(so-called rapidity reversal)
$\tilde{n}(\mathbf{r},t)\leftrightarrow\,-\,n(\mathbf{r},-t)$
for vanishing (symmetry breaking) source $h$.

Using standard techniques known from equilibrium~\cite{BREZIN_5}, it
is possible to calculate size-dependent universal scaling functions as well as
the involved critical exponents. We consider DP in a finite
cubic geometry of linear size~$L$ with periodic boundary conditions and expand
$n$ and $\tilde{n}$ in complex exponential plane waves, e.g.
\begin{equation}
n(\mathbf{r},t)=\sum_{\mathbf{q}}\mathrm{e}^{i\mathbf{q\cdot r}}
n(\mathbf{q},t)\, .
\label{eq:plane_waves}
\end{equation}
Each component of the wavevector takes
only discrete values, precisely multiples of $2\pi/L$ including zero.
Following \cite{JANSSEN_9}, a dynamic free energy functional 
$\Sigma\lbrack{\tilde{\Phi},\Phi]}$ for the $\mathbf{q}=\mathbf{0}$ 
mode is constructed by decomposing the critical homogenous modes 
$\tilde{\Phi}(t)$, $\Phi(t)$ from their orthogonal non-critical 
complements $\tilde{\Psi}(\mathbf{r},t)$, $\Psi(\mathbf{r},t)$, e.g.
\begin{equation}
n(\mathbf{r},t)=\Phi(t)+\Psi(\mathbf{r},t) 
\label{eq:decomposing}
\end{equation}
with $\Phi(t)=L^{-d}\int\mathrm{d}^{d}r \,n(\mathbf{r},t)$. This leads to a
decomposition of the response functional $\mathcal{J}=\mathcal{J}_{0}%
+\mathcal{J}_{1}$ with
\begin{equation}
\mathcal{J}_{0}=\lambda L^{d}\int\mathrm{d}t\,\Big\{\tilde{\Phi}%
\Big(\lambda^{-1}\partial_{t}+\tau+\frac{g}{2}\bigl(\Phi-{\tilde{\Phi}%
}\bigr)\Big)\Phi-h{\tilde{\Phi}}\Big\} . 
\label{eq:J_PHI}
\end{equation}
Now, $\tilde{\Psi}$ and $\Psi$ are eliminated
by a functional integration
\begin{equation}
\mathrm{e}^{-\Sigma [\tilde{\Phi},\Phi]}=
\mathrm{e}^{-{\mathcal J}_0 [\tilde{\Phi},\Phi]} \;
\int\mathcal{D}[\tilde{\Psi},\Psi] \,
\mathrm{e}^{-{\mathcal J}_1 [\tilde{\Psi},\Psi;\tilde{\Phi},\Phi]} .
\label{eq:func_int}
\end{equation}
The part $\mathcal{J}_{1}$ contributes to the leading scaling
behavior for $d\leq d_{\mathrm{c}}$~\cite{JANSSEN_9}. 
This will be revisited in a successional publication~\cite{JaLu04}. 
Here, we consider the mean field
regime ($d>d_{\mathrm{c}}$) where $\mathcal{J}_{1}$ provides, 
besides the shift of the control parameter from its mean 
field value ($\tau_{c}^{\mathrm{mf}}=0$) 
to $\tau_{c,L}=\tau_{c,\infty}+O(L^{2-d})<0$,
corrections to the leading asymptotic scaling behavior. 
Hence, we neglect $\mathcal{J}_{1}$ in the following 
but include the shift of the critical point to the 
infinite size value $\tau_{c}=\tau_{c,\infty}$ by the redefinition 
$\tau\rightarrow\tau-\tau_{c}$.
Correlation functions of the order parameter $\Phi$, 
$\langle \prod_{\alpha=1}^{k}\Phi(t_{\alpha})\rangle=
G_{k}(\{t_{\alpha}\},{\tau},h,L^{d},\lambda,g)$, 
can be derived from path integrals with weight $\exp(-\Sigma)$.
In that way, Eq.~(\ref{eq:J_PHI}) and simple dimensional scaling 
leads to the  parameter reduction in the correlation function
\begin{align}
&  G_{k}(\{t_{\alpha} \},{\tau},h,L^{d},\lambda,g)=
g^{-k}F_{k}(\{\lambda t_{\alpha}\},{\tau},gh,L^{d}/g^{2})\nonumber\\
&  =L^{-kd/2}f_{k}\bigl(\{L^{-d/2}g\lambda t_{\alpha} \},L^{d/2}{\tau}
/g,L^{d}h/g\bigr). 
\label{eq:KorrFu}
\end{align}
Therefore, it is convenient to define
\begin{equation}
\varphi(s)=L^{d/2}\Phi(t)\quad{\mathrm{and}}\quad s=gL^{-d/2}\lambda\,t\,.
\label{eq:def_phi}
\end{equation}
Then Eq.\thinspace(\ref{eq:J_PHI}) yields
\begin{equation}
\Sigma\approx\mathcal{J}_{0}=\int\mathrm{d}s\Big\{\tilde{\varphi}
\Big(\partial_{s}+T+\frac{1}{2}\bigl(\varphi-{\tilde{\varphi}}
\bigr)\Big)\varphi-H{\tilde{\varphi}}\Big\} 
\label{eq:J_PHI_ab}
\end{equation}
where the rescaled control parameter and the rescaled source
\begin{equation}
T=g^{-1}L^{d/2}{\tau} \quad {\mathrm{and}}
\quad H=g^{-1}L^{d}h
\label{eq:def_P_H}
\end{equation}
are introduced. Note that the whole dynamic functional depends on the rescaled
parameters $T$ and $H$ only. Furthermore, the rescaled parameters contain the
irrelevant parameter~$g$ in a dangerous, i.e., singular way.

The rescaled parameters [Eq.\thinspace(\ref{eq:def_P_H})] 
and Eq.\thinspace(\ref{eq:KorrFu}) already contain a
non-trivial result: As usual for critical phenomena, physical quantities of
interest are described in terms of generalized homogenous functions. For
example, the steady state order parameter $n=\langle \Phi \rangle$
and the steady state correlation length~$\xi$
obey for all $l>0$ 
the scaling forms (despite of non-universal metric factors) 
\begin{eqnarray}
\label{eq:scal_order_par}   
n & = & l^{\beta/\nu^{\ast}}\,{\tilde{R}}
(\tau l^{-1/\nu^{\ast}},h l^{-\Delta/\nu^{\ast}},L l), \\
\label{eq:scal_corr}   
\xi & = & l^{-1}\,  {\tilde \Xi}
(\tau l^{-1/\nu^{\ast}},h l^{-\Delta/\nu^{\ast}},L l) ,
\end{eqnarray}
with the universal functions ${\tilde{R}}$ and ${\tilde{\Xi}}$
(analogous scaling
functions are known from equilibrium~\cite{BINDER_3,BRANKOV_2}).
Usual FSS forms involve the correlation length exponent $\nu$ 
whereas the above modified scaling forms contain the so far 
unknown exponent~$\nu^{\ast}$. 
For $d>d_{\mathrm{c}}$, the order parameter exponent $\beta$ and the 
field exponent~$\Delta$ (often called gap exponent in equilibrium) equal their mean 
field values $\beta=1$ and $\Delta=2$. 
Comparing Eq.\,(\ref{eq:KorrFu}) and 
Eqs.\thinspace(\ref{eq:def_P_H}) to the scaling forms 
Eq.\thinspace(\ref{eq:scal_order_par}) and 
Eq.\thinspace(\ref{eq:scal_corr})
for $l=L^{-1}$ yields the FSS exponent for periodic
boundary conditions
\begin{equation}
\nu^{\ast}=\frac{2}{d}. \label{eq:nu_star_pbc}%
\end{equation}
Note that $\nu^{\ast}$ depends on the spatial dimension in 
contrast to the exponents $\beta$ and $\Delta$. 
More important, $\nu^{\ast}$ differs from the known mean
field value of the correlation length exponent $\nu=1/2$ for $d>d_{\mathrm{c}%
}=4$. Thus, the FSS forms are not controlled by the ratio
$L/\xi_{\infty}\propto L\left\vert \tau\right\vert ^{\nu}$ but by $L\left\vert
\tau\right\vert ^{\nu^{\ast}}$. This scaling anomaly occurs within the mean
field regime only and is regarded as breakdown of FSS.
Furthermore, it can be interpreted as an appearance of an additional length
scale, termed thermodynamic length scale $l_{\infty}$, which diverges as
$l_{\infty}\propto\left\vert \tau\right\vert ^{-\nu^{\ast}}$~\cite{BINDER_3,BRANKOV_2}. 
Similar to equilibrium, this length scale coincides with 
$\xi_{\infty}$ below $d_{\mathrm{c}}$, including $\nu=\nu^{\ast}$. 
Thus, the exponent $\nu^{\ast}$ 
fulfills the hyperscaling relation 
$\nu^{\ast} d = 2 \beta + \gamma^{\prime} $ in all dimensions.

Additionally to the critical exponent $\nu^{\ast}$ is it even possible to
derive universal scaling functions, e.g.~${\tilde{R}}(0,x,1)$. The dynamic
functional Eq.\thinspace(\ref{eq:J_PHI_ab}) corresponds to the following
Fokker-Planck equation~\cite{JANSSEN_9}
\begin{equation}
\partial_{s} P(\varphi,s) =
\Big\{
\partial_{\varphi}[(T+\frac{\varphi}{2})\varphi -H ]^{\vphantom{X}}
+ \partial^2_{\varphi} \frac{\varphi}{2}
\Big\}  
P(\varphi,s) .
\label{eq:fokker_planck}
\end{equation}
The stationary solution $P_{0}(\varphi)$
\begin{equation}
P_{0}(\varphi)=C\varphi^{2H-1}\mathrm{e}^{-(2T+\varphi/2)\varphi},
\label{eq:fokker_planck_stat}%
\end{equation}
can be normalized by an appropriate finite factor~$C(H,T)$ for $H>0$. 
Straightforward calculations yield the moments at bulk criticality ($T=0$)
\begin{equation}
\langle\varphi^{k}\rangle=2^{k/2} \;
{\Gamma(H+k/2)} \; / \; {\Gamma(H)}.
\label{eq:ord_par_moments}
\end{equation}
Thus, the universal FSS functions of the order
parameter $n=\langle\Phi\rangle= L^{-d/2} \langle\varphi\rangle$, 
of the order parameter fluctuations 
$\Delta n=L^d(\langle\Phi^{2}\rangle-\langle\Phi\rangle^{2})=
\langle\varphi^{2}\rangle-\langle\varphi\rangle^{2}$ 
as well as of the ratios
$V=\langle\Phi^{2}\rangle/\langle\Phi\rangle^{2}-1$, 
$S=1-\langle\Phi^{3}\rangle/(3\langle\Phi\rangle\langle\Phi^{2}\rangle)$,
$Q=1-\langle\Phi^{4}\rangle/(3\langle\Phi^{2}\rangle^{2})$ are given by
\begin{align}
n &  = \sqrt{\frac{2}{L^d}} \, \frac{\Gamma(\frac{x+1}{2})}{\Gamma(\frac{x}{2})}
=L^{-d/2}\left\{
\begin{array}
[c]{ll}
\sqrt{x}, & x\rightarrow\infty\\
\sqrt{{\pi/2}}\,x, & x\rightarrow0,
\end{array}
\right.  
\label{eq:uni_scal_forms_pbc_n}\\
\Delta n & =x-2\frac{\Gamma(\frac{x+1}{2})^{2}}{\Gamma(\frac{x}{2})^{2}}
=\left\{
\begin{array}
[c]{ll}
1/2, & x\rightarrow\infty\\
x, & x\rightarrow0,
\end{array}
\right. \\
V &  =  \frac{x\Gamma(\frac{x}{2})^{2}}{2\Gamma(\frac{x+1}{2})^{2}}-1=\left\{
\begin{array}
[c]{ll}
1/2x,  & x\rightarrow\infty\\
2/\pi x, & x\rightarrow0,
\end{array}
\right.  \\
S &  =\frac{2}{3}\Big(1-\frac{1}{2x}\Big),
\quad\quad
Q   =\frac{2}{3}\Big(1-\frac{1}{x}\Big),
\label{eq:uni_scal_forms_pbc_SQ}
\end{align}
with the scaling argument $x=2H = 2 h L^{d} / g$. 
In contrast to equilibrium, the ratios $V$, $S$, $Q$ 
are not finite at the critical point ($x\to 0$).
This reflects the different nature of the zero order
parameter phase in equilibrium and in absorbing phase
transitions.
A ratio that remains finite at criticality is obtained via
\begin{equation}
U(x)=\frac{2-3S(x)}{2-3Q(x)}=
\frac{\langle\Phi^{2}\rangle\langle\Phi^{3}\rangle-\langle\Phi\rangle\langle\Phi^{2}
\rangle^{2}}{\langle\Phi\rangle\langle\Phi^{4}\rangle-\langle\Phi\rangle
\langle\Phi^{2}\rangle^{2}}=\frac{1}{2}.
\label{eq:def_ratio_U}
\end{equation}
We expect that this ratio is as useful for absorbing phase transition
as the Binder cumulant~$Q$ is for equilibrium, i.e.,
its value at criticality characterizes the universality class.
Preliminary numerical investigations below $d_{\mathrm{c}}$ yield
$U_{d=1}=0.833$, $U_{d=2}=0.704$, and $U_{d=3}=0.61$ for
$x\to 0$~\cite{JaLu04}.

\begin{figure}[t]
\includegraphics[width=8.3cm,angle=0]{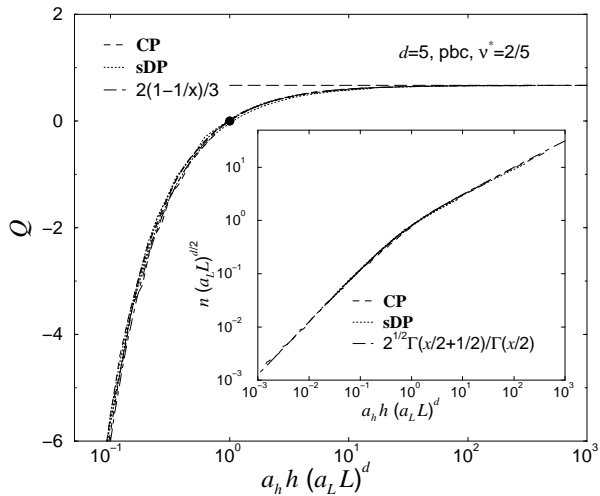} 
\caption{ The universal
order parameter scaling function ${\tilde R}_{\mathrm{pbc}}(0,x,1)$ (inset)
and the universal fourth order ratio scaling function ${\tilde
Q}_{\mathrm{pbc}}(0,x,1)$ as a function of the rescaled field $a_{h} h
(a_{L}L)^{d}$ at criticality for $d>d_{\mathrm{c}}$. The analytically obtained
scaling functions are in perfect
agreement with numerical data of the five-dimensional contact process (CP,
implemented on simple cubic lattices of size $L=4,8,16$, 
$\lambda_{\mathrm{c}}=1.13846(11)$) 
and of the five-dimensional site-directed percolation process
(sDP, implemented via the Domany-Kinzel
automaton~\cite{DOMANY_1} on lattices of a generalized bcc-like 
structure~\cite{LUEB_28}
of linear size $L=8,16,32$,
$p_{\mathrm{c}}=0.0359725(2)$~\cite{GRASSBERGER_P2004}). 
Note that the numerical data already belongs to the
asymptotic scaling regime.
In case of the numerical data, non-universal
metric factors $a_h$ and $a_L$ have been introduced
in order to norm the universal scaling functions,
i.e., ${\tilde R}_{\mathrm{pbc}}(0,0,1)=1$ for the order parameter
and ${\tilde Q}_{\mathrm{pbc}}(0,1,1)=0$ (bold circle)
for the ratio $Q$ (see~\cite{LUEB_28,LUEB_23} for details).}
\label{fig:dp_fss_5d_pbc}%
\end{figure}

The order parameter~$n$ and the ratio~$Q$ are shown in 
Fig.\,\ref{fig:dp_fss_5d_pbc}. 
Additionally to the above derived universal 
scaling functions we plot corresponding
simulation results of the five-dimensional contact process (CP) as well as of
the five-dimensional site-directed percolation process (sDP). Both models 
belong to the DP
universality class (see~\cite{HINRICHSEN_1} and references therein). In
contrast to conventional equilibrium simulation techniques, no steady state
finite-size quantities are available for absorbing phase transitions at zero
field. Close to the transition point, the systems will be soon trapped in the
absorbing state without chance of escape. As recently pointed out
in~\cite{LUEB_23}, the natural way to circumvent these difficulties is to
perform simulations in non-zero field at criticality. Thus both, the
analytical results as well as the numerical simulations reflect that
well-defined steady state quantities exist close to the critical 
point for $h>0$ only.
As can be seen in Fig.\,\ref{fig:dp_fss_5d_pbc}, the data of the lattice
models obey the modified FSS forms and the obtained scaling curves 
are in perfect agreement with the results of the continuum theory. 
A comment is worth making:~In order to reach numerically the asymptotic
scaling regime, the considered system sizes~$L$ have to exceed 
all intrinsic non-universal length scales $L_0$, i.e., 
$L \gg L_0$.
The convincing agreement between the numerical and the field 
theoretical results indicates that $L_0$ is sufficiently
small for the quantities 
Eqs.\,(\ref{eq:uni_scal_forms_pbc_n}-\ref{eq:uni_scal_forms_pbc_SQ}).

In contrast to e.g.~$Q(x)$, the ratio $U$ exhibits a different
behavior.
To be precise, the leading order of $U$ is no
longer a function of the scaling argument $x=2 h L^{d}/g$
within the mean field regime.
Therefore, non-universal corrections to scaling become
dominant.
Analytically, non-universal corrections to $U=1/2$ are 
obtained by incorporating the shift of the critical value
($\tau \rightarrow \tau - O(L^{(4-d)/2})$).
The results are confirmed numerically and will be published
in a forthcoming paper~\cite{JaLu04}.
Again, if the universal leading order of the 
order parameter moments $\langle \Phi^k \rangle$ is 
cancelled for ratios such as~$U$, a non-universal behavior occurs.
Thus $U$ is an appropriate quantity to investigate 
the relevance of corrections to scaling for $d>d_{\mathrm{c}}$.
In summary, a convincing agreement is observed between the 
lowest mode finite-size analysis and corresponding numerical results.
This is in contrast to the situation in equilibrium where it
is known that the simplest lowest mode approach fails to describe
the scaling behavior~\cite{CHEN_3}.


We would like to thank P.~Grassberger for communicating his results prior to
publication. S.~L{\"u}beck thanks A.~Hucht for numerous and fruitful
discussions.


\end{document}